\documentstyle[preprint,aps]{revtex}
\begin{document}
\draft
\title {  Sharp Photoemission Spactra in the  Quantum Antiferromagnet}
\author{ S Sorella}
\address{INFM, International School for Advanced Study, Via Beirut 4, 34013
Trieste, Italy\\
 Instituto de Ciencia de Materiales, Cantoblanco, 28049, Madrid, Spain }
\date{\today}
\maketitle
\begin{abstract}
The low energy
 photoemission spectra in quantum antiferromagnets
are studied by  using
several approximation-free calculations  and rigorous  theorems.
The   important and measurable property found is that
the  hole eigenstates with momenta differing by the antiferromagnetic
wavevector
$Q$ are equivalent  and degenerate in energy. However the corresponding
eigenstates  differs by the presence or the absence of a well defined
quasiparticle  corresponding to a singular -zero energy- magnon, carrying spin
one and momentum $Q$. This difference between the two eigenstates   affects
dramatically the spectral weight as a function of  the scattered momentum,
since
a sharp effect  at the surface of the magnetic Brillouin zone is
predicted, in apparent agreement with recent experimental data.
\end{abstract}
%%%%%%%%%%%%%%%%%%%%%%%%%%%%%%%%%%%%%%%%%%%%%%%%%%%%%%%%%%%%%%%%
\narrowtext
%%%%%%%%%%%%%%%%%%%%%%%%%%%%%%%%%%%%%%%%%%%%%%%%%%%%%%%%%%%%%%%%%%%%%%%%%%%
\section{Introduction}
Soon after the discovery of high-$T_c$ materials it was
proposed that, this kind of interesting phenomenon , could be
explained by the strong antiferromagnetic fluctuations characterizing the
phase at small hole doping.
The fact that in this regime the materials are close to a Mott-Hubbard
transition have stimulated an
intense scientific activity on the study of simple one band models , like
the Hubbard  model and the $t-J$ model, whose apparent simplicity is very
attractive.
However these models  represent a still unsolved many-body problem in
condensed matter physics, as it is still questioned whether they can have
a superconducting ground state in two spatial dimensions.

The so called $t$-$J$ model is defined by the following hamiltonian:
\begin{equation}\label{tj}
H=-t\sum_{<i,j>,\sigma} (c^\dagger _{i\sigma}c_{j\sigma} + h.c. ) +
J \sum_{<i,j>} ({\bf S}_i \cdot {\bf S}_j - {1 \over 4} n_i n_j ).
\end{equation}
where the constraint of no double occupancy is understood.
Here $c_i^\dagger$($c_i$) creates (annihilates) an electron at site $i$,
$n_i=\sum \limits_\sigma n_{i \sigma}$ is the corresponding  density
operator  with $n_{i\sigma}=c^\dagger _{i\sigma} c_{i\sigma}$,
the symbol $<i,j>$ means summation over nearest neighbors, $J$ is
the super-exchange coupling  and finally
the spin density operator ${\bf S}_i$ is defined by the Pauli matrices ${\bf
\vec \sigma} $: ${\bf S}_i =
\sum\limits_{\sigma,\sigma^\prime} c^\dagger_{i,\sigma} {{\bf \vec
\sigma}_{\sigma,\sigma^\prime} \over 2} c_{i,\sigma^\prime}$.
Hereafter  periodic boundary conditions are assumed on a lattice containing
$L$ sites with $N_h$ holes ($L-N_h$ electrons).
In these paper I will mainly focus on the property of a single hole $N_h=1$,
which was a subject of a vast  literature starting from Nagaoka and
Brinkmann and Rice in the early seventies. This problem is also important
because the property  of a single hole in a quantum antiferromagnet are
experimentally accessible  by angle resolved photoemission experiments that
quite recently \cite{wells,aebi} have reached an high level of resolution and
reproducibility especially in two dimensions.
In fact , in absence of hole doping, it is widely accepted that
 the Copper-oxide planes are well described by an effective Heisenberg model
( the term proportional to $J$ in \ref{tj}) with long range
antiferromagnetic order in two dimensions \cite{nelson}. When an electron
is photoemitted , a single hole is free to propagate in the Copper-Oxide
plane, by hopping from Copper site to Copper site with matrix elements
described by the $t$ term in Eq.(\ref{tj}). It is reasonable to assume that
interaction with the $p$ Oxygen bands and other effects cannot lead to
a qualitative change of the model provided the  Coulomb repulsion
remains large to forbid double occupied sites and properly define the position
of the hole in the lattice.
In fact   by
taking the large $U$ limit of the Hubbard model, the lower Hubbard band is
correctly described by the $t-J$  model, which represents therefore a realistic
model  for  the interpretation of  photoemission experiments at
low energies.

One of the most important property is to understand whether a single hole
can propagate as a free particle as in a conventional band insulator.
In this case if an electron is
extracted from the lower band it will remains for infinite
time   with the  given hole momentum $k$.
Within the quasiparticle picture it is clear that interaction can
provide a finite lifetime, but since in an insulator all the possible
states in the lower band are occupied there is no phase space for decaying
processes at least for the lowest possible excitations. As in the Landau theory
for Fermi  liquids it is then possible to define the so called quasiparticle
weight $Z_p$  which measures
the distance between
 the many body state $|p>$ of a single
hole and the  corresponding ground state of the insulator $|H>$ in which a hole
has been  created with momentum $p$ and spin $\sigma$, $c_{p,\sigma} |H>$. The
quasiparticle weight is then given by:
\begin{equation} \label{zeta}
Z_p = |<p| c_{p,\sigma} |H>|^2
\end{equation}
For a band insulator with no interaction $Z_p$ is one by definition, whereas
if  a quasiparticle picture is well defined $Z_p$  remains finite in the
infinite   volume limit. A key question is to understand whether in two spatial
dimension  a quasiparticle theory can be defined for an insulator or if
this quantity vanishes with a power law as in one dimension\cite{parola} In
particular in an antiferromagnet it is crucial to understand if the scattering
with the underlying low energy modes-the spin wave excitations-
can lead to some anomalous effect.

By the Lehman decomposition of the one particle Greens function $G_p(\omega)$
the finiteness of the  quasiparticle weight  $Z_p$ determines a $delta-$
function in the spectral weight $A(\omega<0) = -{1 \over \pi} G_p(\omega)$,
which , as mentioned, is experimentally accessible by ARPES.
However it is difficult to distinguish , in an experiment, which is biased by
the resolution of the electronic device, if a true $\delta$ function exists
or if the spectral weight is completely incoherent , i.e. without $\delta$
function contributions.
This $\delta-$ function weight $Z_p$  occurs at an energy $E_p$, which
represents , within the quasiparticle picture, {\em the lowest one hole energy
with given momentum $p$}, and $E_p$ properly describes the energy dispersion
of
the hole.

In two spatial dimensions there is no exact solution of the ``one hole
problem''
but there is a general believe that the basic dynamical properties  of a
single hole in a quantum antiferromagnet are well described in the
paper  by  Kane Lee and Read (KLR) \cite{klr}>. In this work  the problem is
solved using two main approximations: a large   spin limit and a self
consistent
Born approximation  to sum the relevant diagrams contributing to the Greens
function,  accounting for the scattering of the hole with the spin-waves.
Within  these approximations i) the quasiparticle weight is  {\em finite}
, ii) the minimum possible energy $E_p$  is generally at
momentum $p=({\pi\over 2},{\pi \over 2})$ and finally , as a general
property  coming from the long range antiferromagnetic order, iii)
momenta differing by the  wavevector $Q=(\pi,\pi)$ are
equivalent both for the energy and for the quasiparticle weight:
\begin{eqnarray}\label{afproperty}
Z_p&=&Z_{p+Q} \nonumber\\
E_p&=&E_{p+Q}.
\end{eqnarray}
In the original KLR paper there was some speculation
 that $Z_p$ could vanish away from the bottom of the band even within their
approximations. However  it was shown later \cite{horsch,manusakis} that
 $Z_p$ is strictly  finite  for all possible momenta in the Brillouin  zone as
long as $J>0$.

An accurate determination of the single hole dispersion  was
recently determined by Quantum Monte Carlo and is essentially in agreement with
the KLR result.
The minimum of the band at $p=({\pi\over 2},{p\over2})$ is also confirmed
 experimentally.\cite{wells}
For the quasiparticle weight Quantum Monte Carlo is not accurate enough and ,
quite recently, the accuracy of the KLR theory has been questioned since
numerical data obtained with Lanczos exact diagonalization up to 26 sites
clearly showed that $Z_{p+Q} << Z_p$ at least for $p\sim 0$.\cite{antnew}

 I will show in this paper (and prove for $J=2t$) that while
the spectrum satisfies (\ref{afproperty}) ( rigorously
shown in App.~\ref{theorem-sec} using a variational ansaz), the quasiparticle
weight  $Z_p \ne Z_{p+Q}$.\cite{antnew} This effect induces a kind of Fermi
volume discontinuity   which is consistent with recent experiments by Wells.
In this work it was pointed out that  the photoemission experiments of
insulating  antiferromagnetic materials look similar to the corresponding
metallic ones.\cite{wells}   This simple effect is instead a
direct consequence of a  singular magnon excitation  present in a quantum
antiferromagnet  which cost zero  energy
and change the spin by one and the momentum by $Q$.
 In essence this strange and
measurable effect is intimately related to the  non commutativity of the order
parameter $\vec m$ with the total spin $\vec S^2$. In any
mean field theory one usually selects a direction of the order parameter
and in this way the spin is no longer  defined for the approximate ground
state.
In an experiments however the variation of the total spin is measured
since the  outcoming electrons have a definite spin. An approach which is
capable  to work in a basis with definite spin $\vec S^2$ and order parameter
$\vec m^2$ allows to deal correctly with the excitations and to
determine  the sharpness of the photoemission spectra in a
 quantum antiferromagnet.

\section{Formalism}
\label{formalism}
 I consider the  $t-J$ Hamiltonian (\ref{tj})
and follow  the  formalism recently proposed by
Angelucci {\it et al.}.\cite{antnew}.
 The basic advantage  of this approach is the possibility
to simplify  the local constraint of no doubly occupied sites
 by extending  the hamiltonian to an enlarged space, where the
physical subspace, satisfying the constraint, is obtained when
 the eigenvalue of a  spin-operator $\vec L$ is maximum.
 In this representation the $t-J$ Hamiltonian matrix reads\cite{antnew},
\begin{equation} \label{tjant}
H = E_0+ \sum \limits_{R,\tau_\mu} \left[
t\,(f^{\dag}_{R+\tau_\mu}  f_{R}  + {J \over 4} (1-n_R -n_{R+\tau_\mu})
\right]
\chi_{R,R+\tau_\mu}.
\end{equation}
where    $f^{\dag}_R$ and  $f_R$ are canonical spinless fermion operators
creating or annihilating a hole  at site $R$ and commuting with the
corresponding spin operators $\vec S_R$. Moreover $n_R=f^{\dag}_R f_R$
is the hole number at the site $R$, $N_h =\sum\limits_R n_R$ being the total
number of holes,  and  the operator
\begin{equation} \label{defchi}
\chi_{R,R^\prime} = 2 \vec S_R \, \vec S_{R^\prime}  \,+ \, {1\over 2}
\end{equation}
interchanges  two $S=1/2$-spins at sites $R$ and $R^\prime$.
 Hereafter periodic boundary conditions over a finite hyper-cubic box
with  linear size $l$ and number of sites $L=l^d$ is assumed.
The  constant  shift of the energies $E_0= {J d \over 2} ( 2 N_h-L) $,
which  represents the classical contribution of an antiferromagnetic
N\'eel state,  is also omitted.

The hamiltonian (\ref{tjant}) is naturally defined in a space where
there is a single spin for each site and the hole
can be created over each spin  site by $f^{\dag}$.
%In this way the holes  propagate over an underlying   spin background.
However  the spin value at the position of the holes is clearly
unphysical because in the physical Hilbert space the hole does not carry spin.
As anticipated this difficulty can be easily solved because the hamiltonian
leaves invariant the physical subspace of states  where the spins over
   the sites occupied by the holes
are all frozen to $1/2$ (or $-1/2$)  .
In fact the  following pseudo-spin  operator, measuring the total spin
at the positions of the holes,  and
satisfying the algebra of the angular momentum
\begin{equation}
%\vec S \,=\, \sum\limits_R (1-n_R) \vec S_R , \qquad
\vec L \,=\, \sum \limits_R n_R \vec S_R ,
\label{pseudospin}
\end{equation}
commutes  with the Hamiltonian (\ref{tjant}) and the invariant subspace with
maximum spin $L=L_z={N_h\over 2}$, coincides with the physical Hilbert
space of the $t-J$ model, where the two models have the {\em same }
matrix elements , i.e. they identically coincide.

Analogously the physical total spin has to be measured where there are no holes
and within the present formalism it is defined by:
\begin{equation}
\vec S \,=\, \sum\limits_R (1-n_R) \vec S_R , \qquad
\label{spin}
\end{equation}
Even this operator commutes with the hamiltonian and
obviously with the pseudo-spin $\vec L$.
In the following sections  the total spin
\begin{equation} \label{spintot}
\vec S^{tot}=\vec S + \vec L
\end{equation}
will  be introduced too.

 Any eigenstate $|\psi>$ of the hamiltonian
 can be labeled  by six quantum numbers:
\begin{itemize}
\item The number of holes $N_h$
\item The value of the physical spin $S$ and its component on the z-axis $S_z$
\item The value of the pseudo-spin $\bar L$  and  its component on the
  z-axis $L_z$
\item The total lattice momentum $p$ of the state.
\end{itemize}
and the physical sector is obtained when $\bar L= L_z={N_h \over 2}$.
Correlation functions are expectation value of operators $\hat O$  over these
eigenstates  $<\psi|\hat O |\psi>$. If such an operator commutes with the
total spin $\vec S$ or the pseudo-spin $\vec L$
-as, say, the hamiltonian- the corresponding expectation value
 does not depend on the  spin component
on the z-axis $S_z$ or $L_z$. In particular all the eigenstates differing
by the quantum number $L_z$ or $S_z$ are degenerate.

At the end of this section we remind that a state $|\psi_p>$
 with definite lattice  momentum $p$ is an eigenstate  of
the translation operator $T_R$,
such that $T_R |\psi_p>\,=\, e^{-i p R} |\psi_p>$
The operator $T_R$
is the translation operator that brings the origin $O$
to the lattice point $R$. It is formally defined by the following
relations valid for any $R$ and $R^\prime$:
\begin{eqnarray}\label{translation}
T_R f_{R^{\prime}} T_{-R} \,&=&\, f_{R+R^{\prime}} \nonumber \\
T_R \vec S_{R^{\prime}} T_{-R} \,&=&\, \vec S_{R+R^{\prime}}
\end{eqnarray}
 The lattice momentum $p$  is defined in the Brillouin zone (BZ)
with each component $|p_i| \le \pi$ (the lattice constant is one).
An important region contained in the previous one is the magnetic Brillouin
zone (BZ$^\prime$) defined as the locus of points such that $\gamma_p \ge 0$
where $\gamma_p= { 1 \over 2 d} \sum\limits_{\tau_\mu} e^{ i p \tau_\mu}$.

\section{ Tower of states in a quantum  antiferromagnet}
\label{tower}
In a quantum antiferromagnet the most important
quantities for the correct description of the low energy physics
 are the order
parameter $\vec m$ and the total spin $\vec S$, that, within our formalism,
read:
  \begin{eqnarray}
\vec m &=& { 1 \over L } \sum \limits_R e^{ i Q R} \vec S_R (1-n_R) \nonumber
\\
\vec S &=& \sum\limits_R \vec S_R (1-n_R)
\end{eqnarray}
where $Q$ is the antiferromagnetic wavevector $Q=(\pi,\pi,\cdots)$

The commutation rules of these two operators are well known\cite{siggia}:
\begin{eqnarray}\label{defcomm}
\left[ S_j , m_k \right] &=& i \epsilon_{j,k,l} m_l \nonumber \\
\left[ S_j , S_k \right] &=& i \epsilon_{j,k,l} S_l  \nonumber \\
\left[ m_j, m_k \right]  &=&  { i \over L^2} \epsilon_{j,k,l} S_l
\end{eqnarray}

In particular using the above relations it is easy to obtain the commutation
of $m^{+}=m_x + i m_y$ or $m^{-}= m_x - i m_y$ and $m_z$
with the total spin $\vec S^2$:
\begin{eqnarray} \label{comms2m}
\left[\vec S^2, m^{+} \right] &=& -2 m_z S^{+} + 2 m^{+} (1+S_z) \nonumber\\
\left[\vec S^2, m^{-} \right] &=& 2 m_z S^{-} + 2 m^{-} (1-S_z) \nonumber \\
\left[\vec S^2, m_z \right] &=& m^- S^+ - m^+ S^- + 2 m_z
\end{eqnarray}

 The operator $m^2$ commutes  with $\vec S^2$ and $S_z$. In the
following    it will be assumed that $\vec m^2$ also commutes     with the
hamiltonian  since if long range order sets in $\vec m^2$ becomes a macroscopic
classical variable in the infinite volume limit.\cite{strocchi}  Then we may
classify the eigenstates  in terms  of the four quantum numbers  $\vec
S^2$, $S_z$, $\vec m^2$ and the total lattice momentum of the eigenstate.

Suppose that an eigenstate  $|S,m^2,\vec p>$ is given with total momentum
 $\vec p$, definite spin  $S$ and
maximum azimuthal spin $S_z=S$ , (all the other $S_z$
component can  be easily obtained by a repeated application of the total spin
lowering operator $S^{-}$),  it will be  shown in the following that  the
state $m^{+} |S,m^2,\vec p>$ has  spin $S+1$, $S_z=S+1$, the same  order
parameter $\vec m^2$ but momentum $\vec p+ Q$.
%Since the Hilbert space dimension decreases with increasing spin it is also
%clear  that all these states exhausts the Hilbert space
%of eigenstates  $|S+1,m^2,p+Q>$ of the hamiltonian.

 The change of momentum is easily understood since $ T_R \vec m T_{-R} =
e^{i Q R} \vec  m$, yielding $T_R m^+ |S,m^2,p>= e^{ i (p +Q) R}  m^+
|S,m^2,p>$,
i.e. a state with momentum $\vec p +Q$.  On the other hand,  by hypothesis:  $$
\vec S^2 |S,m^2,p> = S (S+1) |S,m,p> $$  whereas by applying (\ref{comms2m}):
$$ \vec S^2 m^{+} |S,m^2,p> =\left( m^{+} \vec S^2 + \left[ \vec S^2, m^{+}
\right]
 \right)  |S,m^2,p>= (S+1) (S+2) m^{+} |S,m^2,p> ,$$
i.e. $m^{+} |S,m^2,p>$ is a state with definite spin equal to $S+1$.
Analogously $S_z$ is  defined since, by $\left[ S_z, m^{+} \right] =
m^{+}$,  it follows  $S_z m^{+} |S,m^2,p> = (S+1) m^{+} |S,m^2,p>$. Finally
$\vec m^2$ does not change by applying   $m^{+}$ to  the state $|S,m^2,p>$ as
$\left[\vec m^2,m^+ \right]=0$.
\vskip 0.25 truecm

 By the simple assumption that
 $m^2$   commutes with the
hamiltonian for $L \to \infty$, it has been derived that
 all the  different spin sectors are degenerate and the momentum is defined
modulo $Q$.
In a finite system this degeneracy is slightly removed and  in fact the energy
spectrum  as a function of the spin behaves  as a free quantum rotator spectrum
determined by the value of its static susceptibility
$\chi_{ST}$\cite{ziman,bernu}:
$$
E(S) = E_0 + {S (S+1 ) \over 2 L \chi_{ST}}  + o({1\over L}).
$$
Instead  by spin-wave theory the magnon excitations are
characterized by much larger energy costs $\sim {1\over l}$. It seems therefore
exact at least at low enough  energies that the tower of states with different
spin can be considered  degenerate for $L \to \infty$, when $\vec m^2$ becomes
classical.  In a bipartite lattice without holes   the spin is  integer and the
lowest  eigenstates with  $S=n$  can be obtained by applying $n$  times the
operator  $m^{+}$   to the  singlet spectrum of eigenstates  $|S=0,m^2,p>$:
\begin{equation}\label{iterspectrum} |S+n,m^2,p+n Q> \propto
(m^+)^n |S,m^2,p> \end{equation}
In particular, from $2 Q=0$ modulo $2 \pi$,    odd
integer spin have the lowest energy state with momentum $Q$, whereas  even spin
have vanishing  momentum (referenced to the singlet momentum which  is $0$ or
$Q$ according to the parity of $L\over2$ respectively).

In presence of a single hole or an odd number of holes the spin is half odd
integer (for half odd integer spin $s$) and the minimum one is $S={1 \over 2}$.
Consequently the eigenstate  with minimum spin, generator of all the tower of
states in (\ref{iterspectrum}),  has non vanishing spin $S={1\over 2}$. In this
case a  further operator is relevant to generate all the manifold of degenerate
states since by applying $(\vec m \cdot \vec S)$ to an  element
 $|S,m^2,p>$ of the
tower  (\ref{iterspectrum}) one gets a
different state with the same spin (due to the rotationally
invariant expression $(\vec m \cdot \vec S)$, commuting with the total spin)
but momentum $p+Q$:
\begin{equation}
|S,m^2,p+Q> \propto  (\vec m \cdot \vec S) |S,m^2,p>.
\label{psisames}
\end{equation}
In fact,  by using (\ref{defcomm}), the operator
$(\vec m \cdot \vec S)$  commutes  with $m^{+}$ and  obviously
annihilates any singlet state. Hence the expression (\ref{psisames}) defines
non vanishing states only in the half-odd integer spin case where the lowest
spin state   of the tower has $S={1\over 2}$.
The new relation (\ref{psisames}) cannot be iterated as the previous one
(\ref{iterspectrum}), because  the square of the operator $ (\vec S \cdot \vec
m)$ when applied to the generator  state $|S={1\over2},m^2,p>$
behaves as a constant equal to ${(\vec m)^2 \over 4} + o({1\over L})$. In fact
  $(\vec m \cdot \vec S)^2= { 1\over 2} \left[\sum\limits_{i,j} m^i m^j
 ( S^i S^j + S^j S^i )  \right] + { (\vec S)^2 \over 2 L^2 }$ and
 $S^i S^j + S^j S^i= {1 \over 2} \delta_{i,j}$
on a spin-$1\over 2$ state.
The final diagram of the tower of degenerate states (or almost degenerate
states  at finite size) are shown in the following diagram:
\begin{equation}
\begin{array}{ccc}
 \vdots &  \vdots &\vdots \\
|S+1,m^2,p+Q> & \rightleftharpoons & |S+1,m^2,p> \\
              & (\vec S \cdot \vec m) &  \\
\uparrow m^+ &  & \uparrow m^+ \\
|S,m^2,p > & \rightleftharpoons & |S,m^2,p+Q> \\
 & (\vec S \cdot \vec m) &  \\
 \vdots &  \vdots &\vdots \\
|{1\over 2},m^2,p > & \rightleftharpoons & | {1 \over 2},m^2,p+Q>
\end{array}
\label{diagram}
\end{equation}
 We expect therefore that for odd number of holes and fixed spin (and
 maximum azimuthal spin)    all the spectrum is characterized
by  {\em couples } of  eigenstates  with momenta $p$ and $p+Q$
related by the approximate expression (\ref{psisames}).

Indeed for the single hole case it can be rigorously shown (see
App.~\ref{theorem-sec})  that  the lowest  eigenstates with momenta $p$ and
$p+Q$
differ in energy  by terms less than $\sim {1 \over L}$.
Moreover  the relation  (\ref{psisames})  between  such a couple of states is
{\em exact}     within the assumption
that  in any subspace with given spin and momentum  the gap to the
first excited  state scales as  $\sim {1\over l}$ (
one has to excite at least one $k\ne 0,Q$ magnon in this
case), i.e. much larger than the  energy accuracy $\sim {1 \over L}$ of
the state
{}~(\ref{psisames}). In the next section (\ref{exactj2t}) it will be shown that
   Eq.~(\ref{psisames}) is indeed consistent with the exact solution
 obtained for momenta $p=0$ and $p=Q$ at the  supersymmetric point
$J=2t$.\cite{bares}

The general character of the spectrum in a quantum antiferromagnet does not
depend  upon  doping if long range magnetic order
exists with finite momentum $Q$.
However the basic relation (\ref{psisames}) is  expected to be only
approximate at finite doping, because charge excitations with energy cost $\sim
{1\over L}$ are  known to exist (invalidating the assumption of a gap scaling
as
$1\over l$), as it is also clear from the weak coupling  theory in the Hubbard
model.

It is a luck that for a single hole, charge excitations are forbidden by the
requirement of fixed total momentum, and relation (\ref{psisames}) is
asymptotically  exact for  the lowest possible
energy  in each subspace with given momentum. Lowest energy eigenstates
 with momenta differing by $Q$   are  characterized by  quite
different eigenfunctions satisfying relation (\ref{psisames}). In two
dimension momenta $p$ and $p+Q$ are equivalent by  spatial symmetry on the
surface of the magnetic Brillouin zone. Thus one expects
some  discontinuity or at least some singularity  as the momentum of the hole
crosses the magnetic Brillouin zone. This should hold at least  for physical
quantities, like the quasiparticle weight (see  Sec. \ref{qpweight}),
explicitly depending   on  the momentum dependent lowest energy state
$|\psi_p>$.

\section{ Exact integration of the single  hole charge}
\label{intcharge}
The hamiltonian (\ref{tjant})
is translation invariant and  the most general one-hole
state  with  total lattice momentum $-p$ ( hole momentum $p$ ) can be written:
\begin{equation}\label{state}
|\psi_p> \,=\,{1\over \sqrt L } \sum \limits_{R}
e^{i p R} f^{\dag}_{R} T_R  |S>
\end{equation}
where $|S>$ is a pure spin state without holes, i.e. $f_i |S>=0$.

The state $|\psi_p>$ in (\ref{state})
represents the most general one-hole state ($N_h=1$) with
given hole momentum $p$.  The action of any translation invariant
operator $\hat O$ over the state $|\psi_p>$
does not change the momentum of the resulting state $\hat O |\psi_p>$ and is
therefore  equivalent to the action
of an effective spin operator $O^{eff}$ acting on $|S>$ , defined by:
\begin{equation}\label{oeff}
\hat O |\psi_p> = {1\over \sqrt L } \sum \limits_{R}
e^{i p R} f^{\dagger}_{R} T_R  O^{eff} |S>
\end{equation}
For instance consider the kinetic term
of the hamiltonian (\ref{tjant}) along one particular direction $\mu$:
\begin{equation} \label{defkmu}
 K_\mu= \sum\limits_{R^\prime} f^{\dagger}_{R^\prime+\tau_\mu} f_{R^\prime}
\chi_{R^\prime,R^\prime+\tau_\mu}.
\end{equation}
  Using that the state $|S>$ and all the translated ones  $T_{R} |S>$
contain no holes $f_{R^\prime} f^{\dagger}_{R} T_R |S>=
\delta_{R,R^\prime} T_R |S>$,  $K_\mu |\psi_p>$ is easily computed:
$$ K_\mu |\psi_p>= { 1 \over \sqrt{L}} \sum \limits_R f^{\dagger}_{R+\tau_\mu}
 \chi_{R,R+\tau_\mu} e^{i p R} T_R  |S>$$
Then changing $ R+\tau_\mu  \to R $ in the dummy summation and using the
translation operator rule $T_{R-\tau_\mu} = T_R \,T_{-\tau_\mu}$:
$$
K_\mu |\psi_p>= { 1 \over \sqrt{L}} \sum \limits_R f^{\dagger}_{R}
 \, \chi_{R-\tau_\mu,R}\, e^{i p (R-\tau_\mu) } T_R T_{-\tau_\mu}  |S>=
{ 1 \over \sqrt{L}} \sum \limits_R f^{\dagger}_{R}\, e^{i p R } T_R
\,\chi_{O,-\tau_\mu}\, e^{-i p \tau_\mu} T_{-\tau_\mu}  |S>
$$
where  in the latter  equality   $T_{-R} \,\chi_{R-\tau_\mu,R} \,T_R=
\chi_{-\tau_\mu,O}$ comes directly from  (\ref{translation}). Finally,
consistent with Eq. (\ref{oeff}), it follows  that:
\begin{equation}
\label{defkmueff}
K_{\mu}^{eff}=\chi_{O,-\tau_\mu} e^{-i p \tau_\mu} T_{-\tau_\mu}
\end{equation}

Analogously  it is a simple algebra to show that the effective spin
hamiltonian reads:
\begin{equation}\label{heff}
H^{eff}_p \,=\, \sum \limits_{\tau_{\mu}}
 \chi_{O,\tau_{\mu}} ( t e^{i p \tau_{\mu}} T_{\tau_{\mu}}  -J/2) +    H_{SW}
\end{equation}
where  $H_{SW}$ is the translation invariant Heisenberg hamiltonian
$$H_{SW} = {J \over 4}  \sum\limits_{R,\tau_\mu} \chi_{R,R+\tau_\mu} . $$
The total spin $\vec S^{tot}=\sum\limits_R \vec S_R $ of the effective spin
hamiltonian remains unchanged with respect to the old definition
(\ref{spintot}) and  measures the  total spin  in the whole lattice
including the origin site $O$,
whereas the pseudo-spin $\vec L$,  the physical spin $\vec S$ and
the staggered magnetization $\vec m$ turn in:
\begin{eqnarray}
\vec L^{eff} &\to & \vec S_O \nonumber \\
\vec S^{eff} & \to & \vec S^{tot} - \vec S_O \\
\vec m^{eff} & \to & \vec m - { \vec S_O \over L} \label{defmeff}
\end{eqnarray}

  The pseudo-spin and the physical spin  operators  commute with the effective
spin hamiltonian (\ref{heff}), as it is easy to check.  Thus there is a one to
one  correspondence of any eigenstate of the effective  spin hamiltonian with
any
single-hole eigenstate with given momentum of the  extended hamiltonian
(\ref{tjant}).  The identification  of the true  eigenstates of the  $t-J$
model
is in this case trivial, because  the total pseudo-spin $L$ is fixed to
${1\over
2}$  and the extended hamiltonian does not have eigenstates with unphysical
pseudo-spin $L \ne {N_h \over 2}$. This property is valid only for the  one
hole case as it is discussed in  \cite{antnew}.

 In the following it will be  established a
correspondence between the eigenstates of the well known Heisenberg hamiltonian
and the eigenstates of $H_p^{eff}$. To this purpose it is convenient to use the
total spin $ S^{tot}$  and its $z-$ component $S_z^{tot}$ as good quantum
numbers
for the  eigenstates $|S^{tot}, S_z^{tot}>$ of (\ref{heff}).
The total spin of the effective hamiltonian $\vec S^{tot} = \vec L^{eff}+ \vec
S^{eff}$ corresponds to the sum of the physical spin with the pseudospin. It
 commutes with the hamiltonian but does not commute with each component of the
pseudospin $ \vec S_O$.

In order to have a definite total spin $S^{tot}$,
each eigenstate with physical spin $S$ (with all the degenerate $2 S+1$
components) has to combine with the two values of the spin at the  origin,
yielding   eigenstates,   with total spin $S^{tot}=S+{1\over2}$ and  with
$S^{tot}=S-{1\over2}$, by the well known addition relations of angular
momenta.
 After  projecting  each eigenstate $|S^{tot},S_z^{tot}>$  onto  the ones with
definite $S^z_O={1\over 2}$, i.e $|S_O>=({1\over2} +S^z_O ) | S^{tot},
S^{z \, tot}>$ the physical spin of the corresponding hole eigenstates
(\ref{state}) is
  $|S^{tot} \pm 1/2|$.  The physical spin  is thus univocally determined to be
 $S={1\over 2}$  in  the singlet total spin $S^{tot}=0$ subspace.
It is possible to avoid any  ambiguity by restricting  all the following
analysis to this subspace, which in turn  is important for the analysis of
photoemission experiments in stechiometric compounds.

In fact these experiments  determine
the imaginary part of the Greens function, that, at half filling, reads:
\begin{equation}
\label{greenfun}
G(p,t)\,=\, - i <H| c^{\dag}_{p,\sigma}
e^{\displaystyle -i(H -i \delta -E_0) t}
 c_{p,\sigma} |H>
\end{equation}
, i.e. it is obtained by creating a hole over the singlet  antiferromagnetic
state $|H>$.\cite{liebmattis}
 Thus only
the  singlet  subspace of the effective
hamiltonian  is relevant for the calculation of the Greens function  in an
antiferromagnetic insulator.

 In the singlet   subspace we can use
  the  total spin $S^{tot}$ to classify the spin of the
elementary excitations, analogously to what was done in the 1D Heisenberg model
where the spinons have been found to carry spin ${1\over 2}$.\cite{faddev}

The hamiltonian (\ref{heff}) is exact , and  the presence of the translation
operator makes  difficult to use standard approaches  as for the simpler
$H_{SW}$. The recent ansaz\cite{siggia}  proposed by  Shraiman and Siggia
corresponds to a  variational semiclassical solution  of the hamiltonian
(\ref{heff}),
 yielding for
example  the N\'eel state for $H_{SW}$.
However in the one hole case the semiclassical solution cannot be controlled
by the small parameter $1/s$, important to derive the spin-wave limit
for the Heisenberg model.\cite{anderson}

\section{  Quasiparticle weight, Greens function and Current operators}
\label{qpweight}
After the introduction of the effective spin hamiltonian (\ref{heff}) the
Greens function (\ref{greenfun}) is easily expressed  as an expectation value
of a  spin  operator acting on the Heisenberg ground state $|H>$.

The state $|\psi_p> =  c_{p,\sigma}|H>$ is of the form
(\ref{state}), if we choose $|S>=|S_H>$ with:
\begin{equation}
|S_H>\,=\,  n_{\sigma,O} |H>.
\end{equation}
Due to the correspondence of eigenstates between $H_p^{eff}$ and $H$,
we can expand $|S>$ in terms of eigenstates of $H_p^{eff}$ and easily
check that the propagation of $|S_H>$ with the effective Hamiltonian,
 $|S_H>_t = e^{ \displaystyle i H_p^{eff}\,t} |S_H>$, corresponds exactly to
the
propagation of $\psi_p$ with the exact $t-J$ Hamiltonian and the Greens
function immediately follows:
\begin{equation}
G(p,t)\,=\, -{i\over 2}
 < S_H|e^{\displaystyle -i (H_p^{eff} -i\delta) t} |S_H >
\end{equation}
Using that $|S_H>= \sqrt 2 n_{i,\sigma}|H>$, that the commutator
$\left[ H_p^{eff},n_{\sigma,O}\right]$ vanishes and
that $G$ does not depend on $\sigma$,
we get, after Fourier transform $G(p,\omega)=\int \limits_{0}^{\infty} d t
\,\,G(p,t)\,\, e^{ \displaystyle i \omega t} $,
\begin{equation}
G(p,\omega) \,=\, {1\over 2}
 <H| {1 \over  \omega + i \delta  -H_p^{eff} }|H>.
\label{greenp}
\end{equation}
The factor $1\over 2$ is usually omitted in the literature
of strong coupling theories like the $t-J$ model,
 probably for estetic reasons of normalizations. Here  we are interested
to the actual photoemission  delta weight and  we are not allowed to use
misleading normalizations.

Once $H_p^{eff}$ is diagonalized by eigenstates $|i>$ with total
vanishing spin $\vec S^{tot}=\vec S+\vec S_O$ and energies $E_i$ the Greens
function is  obtained by inserting this complete set of eigenstates in
(\ref{greenp}):
\begin{equation} \label{greenlehman}
 G(p,\omega ) = {1\over 2}
\sum \limits_i |<H|p>_i|^2 { 1 \over \omega + i \delta -E_i }
\end{equation}

 A general relation satisfied by the Greens  function
of a single hole in an antiferromagnet directly follows from the property
that  the lowest eigenstate $|p+Q>$ of
$H^{eff}_{p+Q}$  can be written in term of the eigenstate $|p>$ of
$H_p^{eff}$   using relation (\ref{psisames}) and (\ref{defmeff}) valid at low
energy, i.e.    $|p+Q>_i \propto  (\vec S^{eff} \cdot \vec m^{eff} ) |p>_i=
-(\vec S_O \cdot
 \vec m^{eff} ) |p>_i= - (\vec S_O \cdot m) |p>_i$
where in the latter equalities  we have used that $|p>_i$ are singlet states
and we have neglected the $O({1\over L})$ difference between
the physical staggered magnetization $m^{eff}$ (\ref{defmeff})  and
the one  acting over all the sites   $\vec m =\vec m^{eff} + {\vec S_0 \over
L}$.
  A simple normalization is then possible using Eq.~(\ref{normms0}), and by
neglecting $O({1\over L})$ contributions one obtains consistently:
\begin{equation} \label{defs0q}
|p+Q>_i = { 2 \over m } (\vec S_O \cdot \vec m ) |p>_i.
\end{equation}
 for the lowest energy state in each fixed momentum sector.

The imaginary part of the Greens function is experimentally accessible by
angle resolved photoemission experiments (ARPES) and we will refer to it as the
spectral function $A(\omega)$.  In the imaginary part of (\ref{greenlehman})
some
of the $\delta$ function weights  may  remain finite for $L \to \infty$ and
  define the so called quasiparticle weight $Z_p$, which is usually a single
peak located at the bottom of the finite size spectrum\cite{longzhong}.
All the other part of the spectrum merge in a continuum of states for $L
\to \infty$ leading to an incoherent spectral function.

 Using (\ref{greenlehman})
the quasiparticle weight
 is given by:
\begin{equation}
Z_p\,=\, {1\over 2} |<H|S>_p|^2
\label{defzp}
\end{equation}
where $|S>_p$ is the lowest energy singlet state
of the hamiltonian $H^{eff}_p$. Notice that the quasiparticle
weight in a strong coupling theory cannot exceed the value $1\over 2$
since it is obviously bounded by the value of the momentum
distribution $n_p =<H|c^{\dag}_p c_p |H>={1\over2}$.
Notice that using (\ref{defs0q}) $Z_{p+Q}$ can be expressed in the
following  form:
\begin{equation} \label{defzppq}
Z_{p+Q} \,=\,{2 \over m^2}  |<H|  (\vec m \cdot \vec S_O) |S>_p|^2
\end{equation}
This is an asymptotically  exact relation for $L\to \infty$, as long as $m >
0$,
i.e. within the assumption of long range magnetic order.

%Since $H_p^{eff}$ commutes with the spin at the origin $\vec S_O$
%Eq.~(\ref{greenp}) can be equivalently written:
%\begin{equation}\label{greenso}
%G(p,\omega) \,=\, {4\over 3}  <H|\vec S_O \cdot
%{1 \over  \omega   -H_p^{eff}
%}   \vec S_O |H>.
%\end{equation}
%where the normalization factor $4\over 3$ is clearly derived by $\vec S_O
%%\cdot
%\vec S_O = {3\over 4}$ and the spin rotational invariant form has been
%maintained for later convenience. In this latter equivalent expression for
%the Greens function the intermediate states $|p>_i$ in the Lehman
%decomposition  (\ref{greenlehman}) have spin one and the overlap matrix
%elements  $|<H|p>_i|^2$ are replaced by $|<H|\vec S_O |p>_i|^2$. In the large
%spin limit  the latter  expression for the Greens function (\ref{greenso}) is
%more convenient because the spin one excitations are  elementary excitations
%(spin-waves) for the effective spin hamiltonian and, these one magnon
%contributions almost exhausts the sum rule $\int A(\omega)={1\over 2}$.

Another important quantity to study is the   current operator, which is
useful when we calculate the transport properties.
On a  discrete lattice,
the current  operator corresponding to a uniform field
 is defined by\cite{zhangc}:
\begin{equation}
J_\mu = \left[ i e t \sum_{R\sigma}  c^\dagger_{R\sigma}c_{R+\tau_{\mu} \sigma}
+{\rm h.c.} \right] \end{equation}
In terms of spinless fermion hole operators $f_i$ and spin
interchange ones $\chi_{i,j}$ the total current   is given by:
\begin{equation}
J_\mu = \left[ i e t \sum_{R\sigma}   f_R f^{\dagger}_{R+\tau_\mu}
\chi_{R,R+\tau_\mu} +{\rm h.c.} \right]
\end{equation}

We note that $J_\mu= -i e (K_{\mu} - K_{-\mu})$ where $K_\mu$ is defined
in (\ref{defkmu})
,thus when  applied to a state of the form (\ref{state})
 the effective current
operator , acting only on a spin wavefunction, can be written as
\begin{equation}
J_\mu^{eff}= [i e t \chi_{O
\tau_{\mu}} e^{ i p \tau_\mu} T_{\tau_{\mu}}
 +h.c.]
\end{equation}

Following \cite{zhang} the real part of the conductivity in a uniform
field  is characterized by a $\delta$ function proportional to the hole
kinetic energy per site  and a paramagnetic contribution given by the
Kubo formula:
\begin{equation} \label{conductivity}
 \sigma(\omega)_\mu = e^2 {  K_\mu
+  \Lambda_\mu(\omega) \over i (\omega + i \delta) }
\end{equation}
where $K_\mu=t \chi_{0,\tau_\mu}e^{i p \tau_\mu}$ is the kinetic energy in the
 direction  $\mu$ and $\Lambda$ is given by:
\begin{equation}
{\rm Im} \Lambda_\mu ( \omega) =\sum \limits_j |<p| J_{\mu} | p>_j|^2 \delta
(\omega-\omega_j) \end{equation}
where the sum over $j$ indicates the eigenstates with he same momentum $p$
of the single hole ground state, $\omega_j$ being the corresponding energy
excitations referred to the ground state energy.

Of course a factor $1 \over L$ have been dropped out in the previous
expressions for the conductivity
because there is only a single charge allowed to move. At
 small doping the conductivity
 will  be  proportional to the
number of carriers  times the conductivity of the
single hole, i.e. will be  finite in the thermodynamic limit.

\section{ Some  exact eigenstates  for the supersymmetric point}
\label{exactj2t}
At the supersymmetric point for $J=2t$ the kinetic part proportional to $t$
is exactly canceled by the magnetic bonds around the origin in (\ref{heff}). In
fact any eigenstate $|q>$ of the Heisenberg hamiltonian $H_{SW}$
with total  momentum $q=p$  is an exact eigenstate of (\ref{heff}) with
the same hole momentum  $p$ and energy $E_p$.
In fact by  assumption $H_{SW} |p>= E_p |p>$ and  $T_{\tau_\mu} |p>=e^{-i p
\tau_{\mu}} |p>$, and immediately follows that:
\begin{equation}\label{proofj2t}
H_p^{eff} |p>= E_p |p> + t \sum\limits_{\tau_\mu} \chi_{O,\tau_\mu} (e^{i p
\tau_\mu} T_{\tau_\mu} -1) |p> = E_p |p>
\end{equation}

The collection of all possible one hole states with arbitrary hole momentum
has clearly an Hilbert space dimension   $\sim L$ times the dimension
of the Heisenberg model.
 Thus  the Heisenberg eigenstates  with all possible momenta $q$ are not a
complete set for the one hole Hilbert space.  It is remarkable
however  that for $J=2t$ a
considerable  fraction ($\sim {1\over L} $) of all the eigenstates is exactly
known and indeed coincide with the ones of $H_{SW}$.
I have verified numerically up to $26$ site 2D-lattice  that the lowest-energy
one-hole eigenstates with momentum  $p=0$ or $p=Q=(\pi,\pi)$  are of the
previous
type, i.e. particular eigenstates of the   Heisenberg model.
For $p=0$ in {\em any} spatial  dimension the singlet spin  state $|S>$
characterizing the single hole eigenstate (\ref{state}) coincides  with the
true
ground state of the Heisenberg hamiltonian $|H>$, the quasiparticle
weight  is exactly one and exhausts all the spectral weight,
$A_{p=0}= {1\over 2} \delta(\omega -E_{p=0})$.

An interesting feature instead is when the momentum of the
hole  coincides with the antiferromagnetic wavevector $Q=(\pi,\pi,\cdots)$.  In
this case the singlet state $|S>$ characterizing the lowest possible one hole
eigenstate with momentum $Q$ can be obtained using
the first excitation of the Heisenberg model, which has
momentum $Q$  and  is a triplet:
\begin{eqnarray*}
|H,\sigma> ~~&{\rm for }&~~ \sigma=-1,0,1  \\
|H,1>={1 \over \sqrt{2}} S^+ |H,0> &~~& |H,-1>={1 \over \sqrt{2}} S^- |H,0>
\end{eqnarray*}
with energy $ E_{p=0} +{ 1\over L \chi_{ST}} + o({1\over L})$, where
$\chi_{ST}$ is the static spin susceptibility of the Heisenberg model.
The above states can be combined with  the operator $\vec S_O$
which commutes with the hamiltonian, leading to   nine different  spin
eigenstates  $ S_O^j |H_\sigma>$ for $j,\sigma=1,2,3$
with the  same energy and with spin components on the $S=0,1,2$ subspaces. By a
proper linear combination of these states  it is then possible to select an
exact
singlet  eigenstate in the  following way:
$$|S>_Q = \alpha \left(  S^{+}_O  S^- |H,0> - S^{-}_O S^+ |H,0>
+ 2 S^z_O |H,0> \right) $$ where the normalization constant is easily computed,
yielding: $$\alpha^2  \to { 1 \over 3 } + O ({1\over L})$$

After a little  algebra, using that both $|H>$ and $|H,\sigma>$ have definite
momenta,  the overlap of this state with the
Heisenberg ground state is: $$ < H|S>_Q = { 6 \alpha } <H| m^z_Q| H,0> $$,
leading, by means of the Schwartz inequality, to:
\begin{equation} \label{defz0q}
Z_Q = {1\over2} < H|S>_Q^2 \le 18 \alpha^2 <H| (m^z_Q)^2 |H>=2  m^2
\end{equation}
where the latter equality follow from  $|H>$ being a singlet,
yielding $<(m^z)^2 > ={1 \over 3 } m^2$.
%Notice that for $p\ne Q$ one can use
%the lowest triplet state $|H,\sigma>$ with momentum  $-p$ and determine an
%eigenstate of $H_p^{eff}$. A similar analysis
% will lead to a contribution to the spectral weight for this eigenstate
%given by $Z_p \sim { 4 \over L} S(p)$, where $S(p)$ is the magnetic structure
%factor with momentum $p$ ($S(Q) =L m^2$) . This is an important contribution
%$\sim 1 \over L$ for a single eigenstate of $H_p^{eff}$
%considering that the dimension of the Hilbert  space is exponentially large in
%$L$ and by the sum rule all the total contribution to the spectral weight
% sum up to one.

The above  inequality (\ref{defz0q}) is an upper bound for $Z_Q$, but
represents  an exact equality in the infinite size limit.
In fact after  a little algebra the state (\ref{defs0q}) with $|i>=|H>$
saturates  the bound (\ref{defz0q}), and represents
the true  eigenstate with momentum $Q$  within the only assumption that the
finite size gap in each sector of definite total spin and momentum scales as
$1 \over l$  (see App.\ref{theoremcons-subsec}).
This basic assumption has been verified numerically as it is shown in
Fig. 1
% ~\ref{fsizegap}
for $t=0$ and ${J\over t}=2 $ for  various momenta.
In particular in the static limit $t=0$ the spin wave prediction of this gap
 always underestimates the value of the true gap
computed by exact diagonalization on small lattices. In spin wave theory
 this gap  scales as $1\over l$ since
the perturbation induced by the hole affects only by small shift the bare
spin wave energy  excitation $\epsilon_k$. The dispersion $\epsilon_k$ depends
linearly with  momentum $\epsilon_k \sim c |k|$, where $c$ is the spin-wave
 velocity,  and
 the lowest gap is given by $\epsilon_{k={2 \pi \over l}}  \sim {2 \pi c \over
l}$.
A plot of the quasiparticle weight for momentum $Q=(\pi,\pi)$
is shown in Fig.~2
%(\ref{figexact})
 indicating that the finite size estimate
of this quantity is strongly size dependent and without an exact result
$Z_Q=2 m^2$ it would be difficult to decide whether $Z_Q$ remains finite in
the  infinite size limit.
This is in general the case for $J\ne 2t$, where the situation is still unclear
and controversial so far.

The exact determination of the quasiparticle weight, obtained for particular
wavevectors and $J=2t$ has evidenced  a more general property of the Greens
function in an insulator with long range magnetic  order. In fact hole momenta
differing by the antiferromagnetic wavevector $Q$ are in general characterized
by the same energy in the infinite size limit, but  with a quite different
quasiparticle weight. The two one hole eigenstates with momenta $p$ and $p+Q$
inside or outside the magnetic Brillouin zone respectively,  have in fact a
substantially different overlap with the Heisenberg ground state,  because they
are essentially derived   from two {\em orthogonal} states:    the Heisenberg
ground state, and the corresponding
 lowest triplet excitation, respectively.
This property is clearly a general one, valid for all
momenta:
In fact  the relation between the lowest singlet eigenstates of
$H^{eff}_p$ can be written, analogously to Eq.~(\ref{defs0q}),  :
 $$|S>_{p+Q} = { 2 \over m } (\vec S_O \cdot  \vec m) |S>_p$$
At the boundary of the magnetic Brillouin zone, momenta differing by
$Q$ are equivalent by spatial symmetries. Then arbitrary close to this surface
there should be at least a singularity in the quasiparticle weight, because the
lowest energy state change dramatically even with an arbitrary small variation
of momentum. It is reasonable  to expect a jump of the quasiparticle weight
that
according to (\ref{defz0q}) is given approximately  by:

 $${Z(p+Q) \over Z(p)} \sim (2 m)^2 =0.37$$
where we have used that in the Heisenberg antiferromagnet  the numerical
value for the 2D order parameter is $m\simeq 0.305$.\cite{zhongswt}

\section{Conclusions}

In this work I have
discussed a general property of quantum antiferromagnets that can be
detected  by well resolved photoemission experiments that are determined
by the low energy dynamic of a single  hole
in the undoped material.

Contrary to the existing folklore  considering the single hole  problem
as an old, boring and solved issue\cite{klr,siggia}, it is found here
 that at least an effect has been overlooked by the previous
literature, effect that has been recently observed in  the
photoemission experiments by Wells et al. and were independently predicted
in \cite{wurz}.

Such experiments have evidenced the surprising effect that the photoemission
spectra of a quantum antiferromagnet apparently show up the presence of a
Fermi surface as in the corresponding metal at finite doping.

In the present work   this fact is a consequence
of a zero-energy magnon excitation
carrying  the antiferromagnetic-wavevector momentum and having an
infinite lifetime in an antiferromagnet. This magnon is present
in the ground state of the hole for momenta outside the
Brillouin zone , while is absent in the other momentum region. The matrix
elements entering in the spectral weight measured experimentally is jumping
discontinuously along the mentioned surface separating the two momentum
regions.
This effect is just indicating   the presence of this anomalous excitation.

This kind of
excitation is washed out in any mean field treatment since as discussed in the
introduction, in such a case the direction of the order parameter is fixed and
the spin is no more a measurable quantity.

\acknowledgements
I acknowledge useful correspondence of unpublished work
by  A. Parola who pointed out first the effective spin hamiltonian
(\ref{heff}) in 1991. I am   also grateful to  E.  Tosatti, M. Rice,
A. Angelucci,  D. Poilblanc, P. Prelovcek, P. Horsch and M. Muramatsu
 for useful discussions or comments.

\section{Some exact results for $S={1\over2}$}
\label{theorem-sec}
Consider the effective hamiltonian (\ref{heff}) for the single hole
problem:
\begin{eqnarray}
H_p&=&H^t_p+H^J\\
H_p^t&=& t \sum\limits_{\tau_\mu} e^{i p \tau_\mu} \chi_{O,\tau_\mu}
T_{\tau_\mu}\\
 H^J&=& {J \over 2}  \sum \limits_{<R_I,R_J>\ne O } (\chi_{R_i,R_j} -{ 1\over
2})
\label{defhp}
\end{eqnarray}
$H^J$  represents
the translation invariant Heisenberg hamiltonian without all the bonds
connecting the origin of coordinates $O$.
The single hole hamiltonian commutes with the total spin:
\begin{equation}\label{stot}
\vec S^{tot}= \sum \limits_R \vec S_R
\end{equation}
and the spin at the origin $\vec S_O$, whereas the operator
 $\vec S_Q = {1\over \sqrt{L}} \sum\limits_R e^{-i Q R} \vec S_R $
defines long range  order on a state $|\psi>$ if  $<\psi| \vec S_{-Q}
\cdot \vec S_{Q} |\psi> \to m^2 L$ for $L \to \infty$,
where $L$ is the number of sites, and  $m>0$ is the value of the order
parameter.

We want to prove the following theorem:

\noindent {\sl Theorem.}  Given a single hole eigenstate $|p>$ of the
hamiltonian  $H_p$ with $S^{tot}=0$, and energy $E_p$, assuming that long
range order exists in the given state for momenta $Q=(\pi,\pi,\dots)$,
corresponding to antiferromagnetic long range order, then the triplet states:
\begin{equation}\label{deftri}
|\psi^j>=S_Q^j |p> ~~~ {\rm j=1,2,3}
\end{equation}
define a state of the hamiltonian $H_{p+Q}$ with energy expectation value
\begin{equation}\label{thesisq}
E_j={ <\psi^j|H_{p+Q}|\psi^j>\over <\psi^j|\psi^j>}  =E_p + {A
\over L} \end{equation}
where the constant $A$ is given by:
\begin{equation}
A={-4 E_p + 2 <p| \sum\limits_{\tau_\mu} T_{\tau_\mu} e^{i p\tau_\mu} |p>
  \over m^2 L} \to 4 {e_p \over m^2}
\label{defalim}
\end{equation}
where $-e_p< 0$ is the ground state energy per site of the Heisenberg model
obviously independent of $p$. The latter limit is easily obtained, from the
definition (\ref{defalim}) since $T_{\tau_\mu}$ has all eigenvalues  bounded by
one.

\noindent {\sl Proof.}
Since $|p>$ is a singlet $<\psi^j|\psi^j>={1\over 3} L m^2$ is independent of
$j$, as well as $E_j$. Then we consider the following operator:
\begin{equation}
F=\sum\limits_j F^j= {1 \over 2} \sum\limits_j\left[ S_Q^j ( H_{p+Q}  S_Q^j
-S_Q^j H_{p}) +  (S_Q^j H_{p+Q} -H_{p} S_Q^j) S_Q^j \right]
\label{deff}
\end{equation}
By the exact relation:
\begin{equation}
 S_Q^j H_{p+Q} S_Q^j = { 1 \over 2} ( H_p  S_Q^j   S_Q^j
+  S_Q^j   S_Q^j  H_p) + F^j
\end{equation}
, using that , by assumption,  $H_p|p>=E_p |p>$
it easily follows that:
\begin{equation}
{1 \over 3} \sum\limits_j E_j= E_p + { <p| F |p> \over <p| \vec S_Q \cdot \vec
S_{-Q} |p> }
\end{equation}
Moreover  from $T_{\tau_\mu} S_Q = - S_Q T_{\tau_\mu}$ and:
\begin{equation} \label{bascom}
 \left[ S_Q^j , \left[ \chi_{R_i,R_j} , S_Q^j \right] \right]=
 -{8 \over L}  (\vec S_{R_i} \cdot \vec S_{R_j} -S^j_{R_i} S^j_{R_j} )
\end{equation}
yielding   for $S=1/2$:
$$\sum \limits_j \left[ S_Q^j , \left[ \chi_{R_i,R_j} , S_Q^j \right] \right]=
 -{8 \over L}  ( \chi_{R_i,R_j} -1/2),$$
and the following expression for $F$ holds:
$$ F = { t \over 2} \sum \limits_j \sum \limits_{\tau_\mu} e^{i p \tau_\mu}
 \left[ S_Q^j ,\left[ \chi_{O,\tau_\mu},S_Q^j\right]\right] T_{\tau_\mu}
 + { 1\over 2} \left[ S_Q^j, \left[
H^J, S_Q^j  \right] \right] = -{4\over L} H_p + {2 t \over L}
 \sum\limits_{\tau_\mu} e^{ i p \tau_\mu} T_{\tau_\mu}$$

Finally using the above two relations and the fact that $E_j$ is independent
of $j$   the statement is easily proven (\ref{thesisq}).
\subsection{Consequences and remarks}
\label{theorem-subsec}
 Suppose that the state with momentum $p$ is the lowest energy   state
within the restriction of momenta $p$ and $p+Q$ in the Brillouin zone,
and suppose that this lowest energy $E_p$ is obtained when the  total spin
is minimum $S=0$,
it will be shown
 that normalized  singlet state defined in terms of
 $\vec m= {1 \over \sqrt{L}} \vec S_Q$:
\begin{equation}
|p+Q>= \alpha_L (\vec S_O \cdot \vec m) |p> \label{effpq}
\end{equation}
has an  energy expectation value $\bar E_{p+Q} =<p+Q| H_{p+Q} |p+Q>$  arbitrary
close to $E_p$ and yielding the following  bounds for the lowest energy
$E_{p+Q}$ with momentum $p+Q$ in the singlet subspace:
\begin{equation} \label{theseppq}
 E_p \le E_{p+Q} \le E_p + { 3 A \over L}
\end{equation}
\vskip 0.5truecm
{\sl Proof.}
 From:
\begin{equation}
 (\vec S_O \cdot \vec m)^2= {1 \over 4} \vec m^2  + { 1\over L} (\vec S_O
\cdot \vec m )   - {1 \over 2 L^2} ( \vec S_O \cdot \vec S^{tot} )
\label{normms0}
\end{equation}
the normalization constant is given by
 $\alpha_L = { 2 \over \displaystyle m}+ O({1\over L})$ (the expectation
value of $(\vec S_O \cdot \vec m )$ on the singlet state $|p>$  can be bounded
using the Schwartz inequality
for fixed component $j$: $<p| (\vec S_O \cdot \vec m ) |p>  = 3 <p| S_O^j
m^j |p> \le {\sqrt{3} \over 2 } m$)
 Consider now the normalized state:
\begin{equation}\label{pipmu}
|j> =   \sqrt{ 3} \alpha_L  S_O^{j} m^j |p>
\end{equation}
This state  has non vanishing overlap  with the corresponding singlet one
$|p+Q>$: $$ a_0^2= |< j| p+Q>|^2 \ge ({\rm Re }< j| p+Q> )^2 = {1 \over
3}+ O({1\over L}).$$
%In fact  $ ({\rm Re }< \mu| p+Q> ) = {1 \over 2} {\alpha_L^2  \over \sqrt{3 L}
%}  <p| \left [(\vec m \cdot \vec S_O)  S_Q^\mu + h.c. \right] |p>$ and  simple
%commutation relations allows to detect the m

 Since the hamiltonian commutes with $\vec S_O$
$<j| H |j> = {<p| S^{j}_Q H  S^{j}_Q |p> \over <p| S^{j}_Q
S^{j}_Q |p>}$ which corresponds exactly to $E_j$  in the
previous theorem.     By applying the theorem
 $$<j| H | j> =  E_p + A/L$$
On the other hand the state $|j>$ can be written as
$|j>= a_0 |p+Q> + |\psi^\prime>$ where the state $\psi^\prime$ has no component
in the singlet and  $ <\psi^{\prime} | \psi^\prime>=1-a_0^2$.  By hypothesis
this
state  has an energy expectation value  $<\psi^{\prime}| H^{eff}_{p+Q}
|\psi^{\prime}>  \over <\psi^{\prime} | \psi^{\prime}>$ higher than  $E_p$, the
minimum possible energy  between the $p$ and $p+Q$ subspaces. Thus using
the conservation of the spin,  it follows  that:
$$E_p+ A/L =< \mu| H | \mu> \ge  a_0^2 \tilde E_{p+Q} + (1-a^2_0) E_p$$
where  $ \tilde E_{p+Q} = <p+Q| H |p+Q>$ is a variational estimate of
$E_{p+Q}$.
Finally it the latter inequality
 gives: $$ E_p \le E_{p+Q} \le \tilde E_{p+Q} \le E_p +3 A /L $$
which concludes this proof.
\vskip 1. truecm

\noindent {\sl Remark.} The theorem is  more generally  valid even when the
lowest energy state  is no more a singlet state provided the energy gain to
the
$S=1$ and $S=2$  vanishes at least as $1\over L$. In fact the trial states
$|j>$ have component only in the $S=0$, $S=1$ and $S=2$ sectors.

\subsection{Exact variational state}
\label{theoremcons-subsec}
It has been shown in the previous section  that the variational state $|p+Q>$,
defined in the singlet subspace  is
arbitrarily close in energy  to the exact lowest energy singlet state
$|\psi_{p+Q}>$ with   momentum $p+Q$.
It is  reasonable to assume  that the finite size gap in each subspace with
definite  spin and momentum is of the order  $\sim 1\over l$ because it is
determined  by an excitation of at least one magnon (remind that the spin wave
excitations are of order  $c |k|$, where $c$ is the spin wave velocity and the
minimum allowed  $|k|>0$ is of order $1 \over l$).
If the above hypothesis is correct as it can be easily verified numerically
(see Fig.~1)
% \ref{fsizegap})
 it is possible to  show rigorously that the variational state $|p+Q>$
is arbitrarily close to the exact eigenstate $\psi_{p+Q}$.
\vskip 0.5 truecm

In fact suppose $|p+Q>= a_0 |\psi_{p+Q}> + |\psi^{\prime}>$, where by
definition
$|\psi^{\prime}> $ is orthogonal to the lowest state with definite
spin and momentum, and thus satisfying by assumption:
\begin{eqnarray*}
 < \psi^{\prime} | \psi^{\prime} > &=& 1-a_0^2 \\
{ < \psi^{\prime} | H |\psi^{\prime} > \over  < \psi^{\prime}| \psi^{\prime} >}
&\ge& E_{p+Q} +B/l
\end{eqnarray*}
Then it easily follows that:
$$ < p+Q| H |p+Q> = a_0^2 E_{p+Q} +(1-a_0^2) (E_{p+Q} + B/l)=E_{p+Q} +
(1-a_0^2) B/l. $$
The previous relation is compatible with the previously stated theorem
only if
$$ 1-a_0^2= <\psi^{\prime} | \psi^{\prime}>  \le { 3 A l \over B L} \to 0  $$
which proves the statement of this section for $d>1$,  as  $L=l^d$.

\begin{figure} \label{fsizegap}
\caption{Plot of the finite size gap to the first singlet excitation
as a function of the inverce lattice side length in 2D for
(a) the static limit $t=0$ , full dots are obtained by numerical
diagonalizations and triangles represent the spin-wave predictions, (b) $J=2t$
and momentum $p=(0,0)$.  In this case also the gap to the lowest $S=3/2$
excitation of the Heisenberg model scaling to zero as $1\over L$ is shown for
comparison (squares), while the  triangles
represent  the corresponding gap for the Heisenberg lattice (without the hole)
in spin-wave approximation.
 (c) $J=2 t$ and momentum $p=(\pi/2,\pi/2)$ corresponding to
the single hole ground state. The triangles are the same points
as in (b). (d) $J=2 t$ and momentum $(\pi,\pi)$.
The triangles are the same points as in (b).}
\end{figure}

\begin{figure} \label{figexact}
\caption{Plot of the quasiparticle weight for momentum $p=(\pi,\pi)$ as a
function of the lattice size for different values of $J\protect\over t$.
The star indicate the expected asymptotic $L\protect\to \infty$
value  (see text) for $J=2t$. }
\end{figure}
\end{document}